\documentclass[prd,showpacs,superscriptaddress,nofootinbib,floatfix,12pt]{revtex4-1} 
\usepackage[utf8]{inputenc}
\usepackage{amsmath,amssymb}
\usepackage{slashed}
\usepackage{graphicx}
\usepackage{subfigure}
\usepackage[colorlinks]{hyperref}
\usepackage[usenames,dvipsnames]{color}
\usepackage{mathrsfs}
\usepackage{color}
\usepackage{ulem}
\usepackage{bbm}               
\usepackage{xspace}				
\usepackage{epstopdf}
\usepackage{pgffor}		
\usepackage{array}
\usepackage{multirow}

\newcommand{\huzhou}{\affiliation{Yangtze Delta Region Institute (Huzhou), University of Electronic Science and Technology of China, Huzhou 313001, China}}
\newcommand{\uestc}{\affiliation{School of Physics, University of Electronic Science and
Technology of China, Chengdu 610054, China}}
\newcommand{\shanxi}{\affiliation{Institute of Theoretical Physics, Shanxi University, Taiyuan, Shanxi 030006, China}}
\newcommand{\ucas}{\affiliation{School of Physical Sciences, University of Chinese Academy of Sciences (UCAS), Beijing 100049, China}}
\newcommand{\kek}{\affiliation{KEK Theory Center, Institute of Particle and Nuclear Studies (IPNS), High Energy Accelerator Research Organization (KEK), 1-1 Oho, Tsukuba, Ibaraki, 305-0801, Japan}}

\newcommand{\SCNT}{\affiliation{Southern Center for Nuclear-Science Theory (SCNT), Institute of Modern Physics, Chinese Academy of Sciences, Huizhou 516000, China}\vspace{0.5cm}}

\begin{document}

\title{Investigation on the bottom analogs $T^-_{bb}$ of $T_{cc}^{+}$}

\author{Yu-Shan Ren}\email{ryskxky@yeah.net}
\huzhou\uestc\shanxi

\author{Guang-Juan Wang}\email{wgj@post.kek.jp, corresponding author}
\kek

\author{Zhi Yang}\email{zhiyang@uestc.edu.cn, corresponding author}
\huzhou\uestc

\author{Jia-Jun Wu}\email{wujiajun@ucas.ac.cn}
\ucas\SCNT

\date{\today}

\begin{abstract}
We investigate the doubly bottom state $T^-_{bb}$ composed of two bottom mesons with $J^P=1^+$. The potentials are obtained  using the one-boson exchange model. 
With the heavy quark flavor symmetry, all the parameters in the model are determined by fitting the experimental data of doubly charmed state $T_{cc}^{+}$ from our previous work. 
Our analysis indicates that the isospin symmetry is well-preserved. We identify a deeply bound  $T^-_{bb}$ state  with quantum numbers $I(J^{P})=0(1^+)$, in contrast to the loosely bound $T_{cc}^{+}$. 
Additionally, we discover a resonant $T^-_{bb}$ state with $I(J^{P})=1(1^+)$. 
Furthermore, our investigation into the $\bar{B}\bar{B}^*$-$\bar{B}^*\bar{B}^*$ coupled channel effect reveals its important impact. The binding energy of the bound   $I(J^{P})=0(1^+)$ states becomes deeper, while  the resonant $T^-_{bb}$ state with $I(J^{P})=1(1^+)$ remains nearly unchanged.

\end{abstract}

\maketitle

\section{Introduction}
The dynamical structure of exotic hadrons, which challenges
the conventional quark-antiquark meson or three-quark baryon interpretation, 
is one of the most valuable subjects in modern hadron physics. 
With the help of experimental observations, an increasing number of new hadrons were intensively investigated, primarily in the heavy-quark sector. 
Numerous theoretical studies have actively discussed these exotic states. 
However, their internal structures and properties are still puzzling, 
see reviews~\cite{Chen:2016qju,Lebed:2016hpi,Guo:2017jvc,Liu:2019zoy,
Olsen:2017bmm,Meng:2022ozq,Chen:2022asf}
and references therein for more details. 

After the observation of the hidden-charmed state $X(3872)$~\cite{Belle:2003nnu}, 
an intriguing question whether there exist similar double-heavy tetraquark states 
arises and has garnered significant interest  from both theoretical and experimental perspectives.
The existence of double-heavy tetraquark $QQ\bar{q}\bar{q}$ 
has been widely discussed by various phenomenological models~\cite{Vijande:2003ki,Janc:2004qn,
Vijande:2006jf,Ebert:2007rn,Zhang:2007mu,Karliner:2017qjm,Eichten:2017ffp,Yu:2019sxx, 
Cheng:2020wxa,Chen:2021tnn}, lattice nonrelativistic QCD~\cite{Francis:2016hui,
Junnarkar:2018twb,Leskovec:2019ioa,Mohanta:2020eed,Hudspith:2023loy} and lattice QCD
~\cite{Bicudo:2015kna,Bicudo:2016ooe,Aoki:2023nzp}. 
Recently, the LHCb collaboration reported the discovery of the doubly-charmed tetraquark 
$T^{+}_{cc}$ state in the $D^0D^0\pi^+$ invariant mass spectrum. This state lies slightly 
below the $D^{0}D^{*+}$ threshold with a mass difference of 
$M_{T_{cc}^{+}}-M_{D^0D^{*+}}=-360\pm40^{+4}_{-0}$ keV and has a narrow width of 
$48\pm2^{+0}_{-14}$ keV using the unitary analysis~\cite{LHCb:2021auc}.
It has distinctive exotic flavor composition of $cc\bar{u}\bar{d}$ and spin-parity 
quantum number $J^P=1^+$, indicating an $S$-wave coupling to $D^{*+}D^{0}$. 
These remarkable properties prompt speculation 
regarding $T^{+}_{cc}$ as a hadronic molecule generated 
by the $DD^{*}$ interaction~\cite{Du:2021zzh,Meng:2021jnw,Chen:2021vhg,Ling:2021bir,
Ren:2021dsi,Feijoo:2021ppq,Yan:2021wdl,Ke:2021rxd,Dong:2021bvy,Huang:2021urd,
Fleming:2021wmk,Hu:2021gdg,Chen:2021cfl,Albaladejo:2021vln,Deng:2021gnb,Agaev:2022ast,
Braaten:2022elw,He:2022rta,Abreu:2022lfy,Wang:2022jop,Wang:2023ovj}. 
Other explanations for $T^{+}_{cc}$ have also been proposed, 
such as  the compact tetraquark~\cite{Agaev:2021vur,Xin:2021wcr,Azizi:2021aib,
Abreu:2022lfy}.  In Ref.~\cite{Wang:2023ovj}, we analyzed the $D^0D^0\pi^+$ invariant 
mass distribution for $T^+_{cc}$ in the $DD^*$ molecular picture. The $DD^*$ interactions were described by the one-boson exchange (OBE) model, where 
the potential between two hadrons is attributed to the exchange of light mesons 
including $\pi$, $\rho$, $\omega$. 
The experimental data were well described in the hadronic molecular picture, supporting $T_{cc}^{+}$ as a $DD^*$ molecule. 
By analyzing the properties of $T_{cc}^{+}$  with the complex scaling method (CSM), 
we found the $T_{cc}^{+}$ is predominantly an isoscalar state.

The bottom analogs of $T_{cc}^{+}$,  $T^-_{bb}$ with quark content $bb\bar{u}\bar{d}$, have not yet been observed experimentally. 
The theoretical study of  $T^-_{bb}$ states  has a
longstanding history, utilizing various methods, such as heavy meson effective field theory \cite{Tornqvist:1993ng,Ohkoda:2012hv}, lattice QCD~\cite{Bicudo:2015kna,Bicudo:2016ooe,Aoki:2023nzp}, quark model~\cite{Barnes:1999hs,Yang:2009zzp,Vijande:2003ki,Vijande:2006jf,Ebert:2007rn,Ebert:2007rn,Meng:2023for}, and QCD sum rule \cite{Du:2012wp,Navarra:2007yw,Agaev:2018khe,Agaev:2020dba}. The $T_{bb}^-$ system is predicted to be a deeply bound  state with $I(J^P) = 0(1^+)$ according to lattice QCD calculation \cite{Bicudo:2015kna,Bicudo:2016ooe,Aoki:2023nzp} and  phenomenological models \cite{Vijande:2003ki,Janc:2004qn,Vijande:2006jf,Ebert:2007rn,Zhang:2007mu,Chen:2021tnn,Deng:2021gnb,Karliner:2017qjm,Eichten:2017ffp,Cheng:2020wxa,Francis:2016hui,Junnarkar:2018twb,Leskovec:2019ioa,Mohanta:2020eed,Hudspith:2023loy,Silvestre-Brac:1993zem,Semay:1994ht,Du:2012wp,Navarra:2007yw,Agaev:2018khe,Agaev:2020dba,Meng:2023for,Wang:2018atz,Yu:2019sxx,Ren:2021dsi,Ke:2021rxd,Ding:2020dio,Dai:2022ulk,Feijoo:2023sfe,Maiani:2022qze}, in contrast to the more loosely bound $T_{cc}^+$. Furthermore, the coupled channel effect between the $\bar B \bar{B}^{*}$ and $\bar B^{*}\bar{B}^{*}$ are expected to be more pronounced than that in the charmed sector due to the small mass difference. These suggest quite different inner structures between the $T_{bb}^-$ and $T_{cc}^+$ states. Additionally, other $\bar B^{(*)}\bar{B}^{(*)}$ bound states exists with quantum number $I(J^P) =1(1^+)$ are also predicted~\cite{Li:2012ss,Liu:2020nil,Manohar:1992nd,Zhao:2021cvg,Zhao:2021cvg,Janc:2004qn,Wang:2018atz,Ding:2020dio,Feijoo:2023sfe,Yu:2019sxx,Sakai:2023syt,Oset:2022rey,Ke:2021rxd}. However, studies within the nonrelativistic quark model~\cite{Barnes:1999hs} indicate that only the $I=0$ $B\bar{B}^{*}$ interaction is sufficiently attractive to support a bound state. Thus, investigating into  $T^-_{bb}$ can not only enrich the spectroscopy of the double-heavy tetraquark states, but also serve to examine the theoretical studies and deepen our understanding of these exotic states.

In this paper, we carry out a comprehensive study of the $T^-_{bb}$ states using the complex scaling method (CSM)~\cite{Myo:2014ypa,Myo:2020rni,Aoyama:2006,moiseyev1998quantum,Myo:2014ypa,Chen:2023eri,Moiseyev:1998gjp} and $T$-matrix pole analysis~\cite{Nakamura:2015rta,Wu:2012md,Wu:2014vma}. The coupled channel effect between $BB^*$ and $B^*B^*$ is also explored.
For these states, the interaction is dominated by the light quarks while the heavy quarks are just spectators. This allows us to explore the intriguing characteristics pertaining to the unknown $T^-_{bb}$ from the observed $T^{+}_{cc}$. Within the heavy quark symmetry, the coupling constants in the study of $T^-_{bb}$ are taken to be the same as those obtained from fitting the $T^{+}_{cc}$ experimental data. This approach has been successfully applied in the heavy-strange mesons~\cite{Yang:2021tvc,Yang:2022vdb}. 

This paper is organized as follows. 
In Section~\ref{sec:formul}, we  provide a brief overview of the methodology for calculating effective potential by one-boson exchange (OBE) model.
Additionally, we present the details of complex scaling method and $T$-matrix pole analysis for identifying the $T^-_{bb}$ states. 
Section~\ref{sec:num} is  dedicated to presenting the numerical results of both single and coupled channel analyses and a discussion of our results. 
Finally, a concise summary is given in Section~\ref{sec:sum}. 

\section{Formulation} \label{sec:formul}
\subsection{The effective potentials}\label{sec:eff}
The interactions between the heavy mesons can be described through the exchange of the $\pi$, $\rho$, $\omega$ mesons. The contributions of $\sigma$ and $\eta$ mesons are ignored because their contributions are tiny~\cite{Li:2012cs,Li:2012ss,Wang:2018jlv}.
Within the heavy quark limit and chiral symmetry, the effective Lagrangians can be written as follows~\cite{Falk:1992cx,Cheng:1992xi,Casalbuoni:1996pg,Ding:2008gr,Sun:2011uh}: 
\begin{eqnarray}
\mathcal{L}_{P^{(*)}P^{(*)}M} 
     &=&-i\frac{2g}{f_\pi}\varepsilon_{\alpha\mu\nu\lambda}
     v^\alpha P^{*\mu}_{b}\partial^\nu M_{ba}{P}^{*\lambda\dag}_{a}\ 
     +\ i \frac{2g}{f_\pi}\varepsilon_{\alpha\mu\nu\lambda} 
     v^\alpha\widetilde{P}^{*\mu\dag}_{a}\partial^\nu{}M_{ab}
     \widetilde{P}^{*\lambda}_{b} \qquad \quad  
    \nonumber\\
    &~&-\frac{2g}{f_\pi}(P_bP^{*\dag}_{a\lambda}+P^{*}_{b\lambda}
    P^{\dag}_{a})\partial^\lambda{} M_{ba}
    \ +\ \frac{2g}{f_\pi}(\widetilde{P}^{*\dag}_{a\lambda}\widetilde{P}_b+ \widetilde{P}^{\dag}_{a}\widetilde{P}^{*}_{b\lambda})\partial^\lambda M_{ab}, 
\label{Lagrangian:p}
\end{eqnarray}
\begin{eqnarray}
\mathcal{L}_{P^{(*)}P^{(*)}V}
     &=& -\sqrt{2}\beta{}g_VP_b v\cdot\hat{\rho}_{ba}P_a^{\dag}
        -2\sqrt{2}\lambda{}g_V \varepsilon_{\lambda\mu\alpha\beta}
        v^\lambda(P^{}_bP^{*\mu\dag}_a +P_b^{*\mu}P^{\dag}_a)
        (\partial^\alpha{}\hat{\rho}^\beta)_{ba}
        \nonumber\\
    &~&+\sqrt{2}\beta g_V\widetilde{P}^{\dag}_a v\cdot\hat{\rho}_{ab} \widetilde{P}_b
       -2\sqrt{2}\lambda{}g_V\varepsilon_{\lambda\mu\alpha\beta}v^\lambda
       (\widetilde{P}^{*\mu\dag}_a\widetilde{P}^{}_b+\widetilde
       {P}^{\dag}_a\widetilde{P}_b^{*\mu})(\partial^\alpha{}\hat{\rho}^\beta)_{ab} 
        \nonumber\\
    &~&+\sqrt{2}\beta{}g_V P_b^{*}\cdot P^{*\dag}_a v\cdot\hat{\rho}_{ba}
       -i2\sqrt{2}\lambda{}g_V P^{*\mu}_b (\partial_\mu{}\hat{\rho}_\nu 
       -\partial_\nu{}\hat{\rho}_\mu)_{ba}P^{*\nu\dag}_a \nonumber\\
    &~&-\sqrt{2}\beta g_V\widetilde{P}^{*\dag}_a\cdot\widetilde{P}_b^{*} v\cdot\hat{\rho}_{ab}
        -i2\sqrt{2}\lambda{}g_V\widetilde{P}^{*\mu\dag}_a(\partial_\mu{} \hat{\rho}_\nu 
        -\partial_\nu{}\hat{\rho}_\mu)_{ab}\widetilde{P}^{*\nu}_b, 
\label{lagrangian:v}
\end{eqnarray}
where $P_a=(B^-,\bar{B}^0,\bar{B}_s^0)$ and $P^*_a=(B^{*-},\bar{B}^{*0},\bar B_s^{*0})$ 
represent the heavy meson fields while $\widetilde{P}_a=(B^+,B^0,B_s^0)$ and
$\widetilde{P}^*_b=(B^{*+},B^{*0},B_s^{*0})$ are the
heavy anti-meson fields.
The exchanged pseudoscalar meson and vector meson matrices, $M$ and
$\hat{\rho}^{\mu}$, are distinctly expressed as
\begin{eqnarray}
M=\left( \begin{array}{ccc}
   {\pi^0\over\sqrt{2}}+{\eta\over\sqrt{6}} & \pi^+  & K^+ \\
   \pi^- & -{\pi^0\over\sqrt{2}}+{\eta\over\sqrt{6}} & K^0 \\
    K^-  & \bar{K}^0                                 & -{2\over\sqrt{6}}\eta \\
\end{array}\right ), \quad
\hat{\rho}^{\mu}=\left(\begin{array}{ccc}
  {\rho^0\over\sqrt{2}}+{\omega\over\sqrt{2}}      & \rho^+ & K^{*+} \\
  \rho^-  & -{\rho^0\over \sqrt{2}}+{\omega \over \sqrt{2}} & K^{*0} \\
  K^{*-}  & \bar{K}^{*0}                                    & \phi   \\
\end{array}\right)^{\mu}.
\end{eqnarray}
The $\bar B^{(*)}\bar{B}^{(*)}$ potentials read, 
\begin{equation}
	{\cal V} = V_{\pi}^{u}+V_{\pi}^{t}+V^{t}_{\rho/\omega}+V^{u}_{\rho/\omega}.
\label{eq:v}
\end{equation}
The explicit forms could be found in the Appendix~\ref{app:vBB}. These potentials are derived assuming point-like particles. 
To account for the finite size of the interacting mesons, ensuring regularization and convergence, it is necessary to introduce a form factor. The final effective potential is given by
\begin{equation}
   {\cal V} = \left(V_{\pi}^{u}+V_{\pi}^{t}+V^{t}_{\rho/\omega}+V^{u}_{\rho/\omega}\right)\left(\frac{\Lambda^{2}}
	{\Lambda^{2}+p_{f}^{2}}\right)^{2}\left(\frac{\Lambda^{2}}
	{\Lambda^{2}+p_{i}^{2}}\right)^{2},\label{eq:ff}
\end{equation}
where $\Lambda$ is the cutoff parameter. 

 In Table \ref{tab:obe}, we summarize the momentum-space effective potentials under the isospin bases to  directly illustrate the relationships between different $I(J^{P})$ states.  Notably, we consider $B^{-}\bar B^{*0}$ and $\bar B^{0}B^{*-}$ as distinct channels in our calculation, each with its own independent but remarkably similar effective potential.
%

\begin{table}[h]
	\caption{The $\bar B^{(*)} \bar B^{(*)}$ interactions within the one-boson-exchange model under the isospin symmetry limit.} 
	\label{tab:obe}
	\begin{ruledtabular}
		\begin{tabular}{c|ccc}
		$I(J^P)$~ 
		    & Wave function 
			& ${\pi}$ 
			& $\rho/\omega$ \tabularnewline \hline
        $0(1^+)$
			& $\frac{1}{\sqrt{2}}(B^{-}\bar B^{*0}-\bar B^{0}B^{*-})$ 
			& $\frac{3}{2}V_{\pi}^{u}$ 
			& $\frac{3}{2}V_{\rho}^{u}(\lambda)-\frac{1}{2}V_{\omega}^{u}(\lambda)-\frac{3}{2}V_{\rho}^{t}(\beta)+\frac{1}{2}V_{\omega}^{t}(\beta)$  \tabularnewline
        $1(1^{+})$ 
                & $\frac{1}{\sqrt{2}}(B^{-}\bar B^{*0}+\bar B^{0}B^{*-})$  
                & $\frac{1}{2}V_{\pi}^{u}$ 
                & $\frac{1}{2}V_{\rho}^{u}(\lambda)+\frac{1}{2}V_{\omega}^{u}(\lambda)+\frac{1}{2}V_{\rho}^{t}(\beta)+\frac{1}{2}V_{\omega}^{t}(\beta)$ \tabularnewline    
        $0(1^+)$
			& $\bar B^{*} \bar B^{*}$
			& $V_{\pi}^{u}-\frac{1}{2}V_{\pi}^{t}$ 
			& $V_{\rho}^{u}(\lambda)+V_{\rho}^{u}(\beta)-\frac{1}{2}[V_{\rho}^{t}(\lambda)-V_{\omega}^{t}(\lambda)+V_{\rho}^{t}(\beta)-V_{\omega}^{t}(\beta)]$ \tabularnewline
        $0(1^+)$ 
                & $\bar B \bar B^{*}$-$\bar B^{*} \bar B^{*}$
                & $\frac{3}{2\sqrt{2}}[V_{\pi}^{u}-V_{\pi}^{t}]$
                & $\frac{1}{\sqrt{2}}[\frac{3}{2}V_{\rho}^{u}(\lambda)-\frac{1}{2}V_{\omega}^{u}(\lambda)-\frac{3}{2}V_{\rho}^{t}(\lambda)+\frac{1}{2}V_{\omega}^{t}(\lambda)]$  
		\end{tabular}
	\end{ruledtabular}
\end{table}

\subsection{Complex scaling method}
\label{sec:csm}

\begin{figure}[htp]
\centering
\begin{tabular}{ccc}
\includegraphics[width=0.5\textwidth]{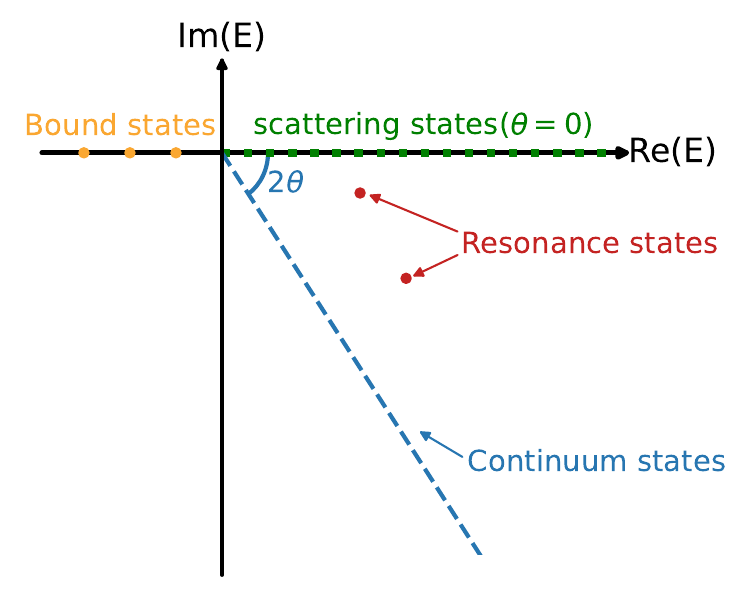}&
\end{tabular}
\caption{The schematic eigenvalue distribution with respect to the complex scaling angle $\theta$ in CSM method.} \label{fig:csmenergy}
\end{figure}

In this subsection, we introduce how  the CSM is used to search for possible bound or resonant states. 
The Schr\"{o}dinger equation in momentum space  is expressed as follows:
\begin{eqnarray}
{\cal H}_0(p)\Phi_i(p)+\int_0^{+\infty} \frac{p^{\prime 2} d p^{\prime}}{(2 \pi)^3} {\cal V}_{ij}\left(p, p^{\prime}\right) \Phi_j\left(p^{\prime}\right)=E\Phi_i(p),
\end{eqnarray}
where ${\cal H}_0=\sqrt{m_{i,1}^{2}+p^{2}}+\sqrt{m_{i,2}^{2}+p^{2}}$ is the kinematic energy with $m_{i,1/2}$ the masses of the constituents in the $i$-th channel. ${\cal V}$ is the effective potential in Eq.~\eqref{eq:ff}.
In CSM, the relative particle coordinate $\boldsymbol{r}$ and its conjugate momentum $\boldsymbol{q}$ are transformed into the complex plane using an operator $U(\theta)$, defined as
\begin{eqnarray}
U(\theta)~:~\boldsymbol{r} \rightarrow \boldsymbol{r} e^{i \theta}, \qquad \boldsymbol{q} \rightarrow \boldsymbol{q} e^{-i \theta}.
\label{eq:CSM}
\end{eqnarray}
Here $\theta$ is a real positive scaling angle. 
Correspondingly, the complex-scaled Schr\"{o}dinger equation could be written as
\begin{eqnarray}
{\cal H}_0({p}e^{-i\theta}) \Phi_i({p}e^{-i\theta})+e^{-3i\theta}\int_{0}^{+\infty} \frac{{p}^{\prime 2} d {p}^{\prime}}{(2 \pi)^3} {\cal V}_{ij}\left({p}e^{-i\theta}, {p}^{\prime}e^{-i\theta}\right) \Phi_j\left({p}^{\prime}e^{-i\theta}\right)=E_{\theta}\Phi_i({p}e^{-i\theta}),\label{csm-eq}
\end{eqnarray}
%
where the eigenvalues become complex and their evolution with respect to the scaling angle $\theta$ is governed by ABC theorem~\cite{Aguilar:1971ve,Balslev:1971vb}. As illustrated in Fig.~\ref{fig:csmenergy}, the eigenvalues for the scattering states lie on the continuum lines that rotated by  $2\theta$ from the real energy axis in the complex energy plane. Each continuum line starts from the threshold energy of the cluster emission. 
The energy eigenvalues of the bound and resonant states are independent of $\theta$ and remain constant  as $\theta$ changes. 
For the resonant pole ($E_{\text{pole}}=E_r-i\frac{\Gamma}{2}$), the scaling angle $\theta$ should be larger than  $\frac{1}{2} {tan}^{-1}(\Gamma/2 E_r)$ to ensure the normalizability of wave functions when integrating  along path $pe^{-i\theta}$. This normalization allows resonant states to be treated similarly to bound states. However, $\theta$ should not be excessively large to alter the damping behavior of the potential at large momenta. More details on CSM can be found in Refs.~\cite{Myo:2014ypa,
Myo:2020rni,Aoyama:2006,moiseyev1998quantum,Myo:2014ypa,Chen:2023eri}. 
\subsection{$T$-matrix pole analysis}\label{sec:tmax}
In this subsection, we briefly introduce the $T$-matrix pole analysis. 
The scattering $T$-matrix describes the dynamical generation of the molecular states via the relativistic Lippmann-Schwinger equation~\cite{Nakamura:2015rta,Wu:2012md,Wu:2014vma}, which is defined as
\begin{eqnarray}
T_{ij}\left(k, k^{\prime} ; E\right)= & \mathcal{V}_{i,j}\left(k, k^{\prime} ; E\right)+\sum_{i^{\prime}} \int_{0}^{\infty} q^{2} d q \mathcal{V}_{i, i^{\prime}}(k, q ; E) G_{i^{\prime}}(q ; E) T_{i^{\prime}, j}\left(q, k^{\prime} ; E\right),
\end{eqnarray}
where $k$, $k^{\prime}$, and  $q$ are the relative momenta in the initial, final, and intermediate states, respectively.  $E$ is the scattering energy. $G_{i^{\prime}}(q ; E)$ is the propagator for  $i^{\prime}$-th channel and is given by:
\begin{eqnarray}
G_{i^{\prime}}(q ; E)=\frac{1}{E-E_{i^{\prime},a}(q)-E_{i^{\prime},b}(q)+i \varepsilon}, 
\end{eqnarray}
where $E_{i^{\prime},a/b}(q)=\sqrt{m_{i^{\prime},a/b}^{2}+q^{2}}$ is the kinematic energies of the intermediate meson $a$/$b$. 
The pole positions of bound states, virtual states, or resonances can be accurately identified by solving the equation $\text{det}[1 - V G(E_{\text{pole}})] = 0$.
If the interaction is attractive and sufficiently strong to generate a bound state, the pole would appear below the two-hadron threshold on the first (physical) Riemann sheets (RS). 
If the strength is insufficient, the pole will appear on the second RS as a virtual state, still below the threshold.  If the interaction is even weaker, the poles are located above the threshold on the second RS corresponding to the resonances.

\section{Results and discussions}\label{sec:num}

The effective potential in Eq.~\eqref{eq:ff} involves two unknown coupling constants $\lambda$ and $\beta$ related to the vector meson exchange. With heavy quark flavor symmetry, we have used the same values for these parameters as determined in our prior work through fitting the experimental $T_{cc}^{+}$ data~\cite{Wang:2023ovj},
\begin{eqnarray}  \label{eq:para08}
&\lambda_{\Lambda=0.8}=0.890/\text{GeV},\quad\beta_{\Lambda=0.8}=0.810; \nonumber\\
&\lambda_{\Lambda=1.0}=0.683/\text{GeV},\quad\beta_{\Lambda=1.0}=0.687;\label{parset} \\
&\lambda_{\Lambda=1.2}=0.587/\text{GeV},\quad\beta_{\Lambda=1.2}=0.550.\nonumber 
\end{eqnarray}

\subsection{$\bar{B}\bar{B}^*$ system}\label{sec:1chan1}
We first study the $\bar{B}\bar{B}^*$ without considering the coupled channel effects with $\bar{B}^*\bar{B}^*$ channel. One bound state and one resonant state are identified with their masses and width summarized in Table \ref{tab:Tbb} and Table \ref{tab:resonance}, respectively.

\begin{figure}[htp]
\centering
\begin{tabular}{ccc}
\includegraphics[width=0.5\textwidth]{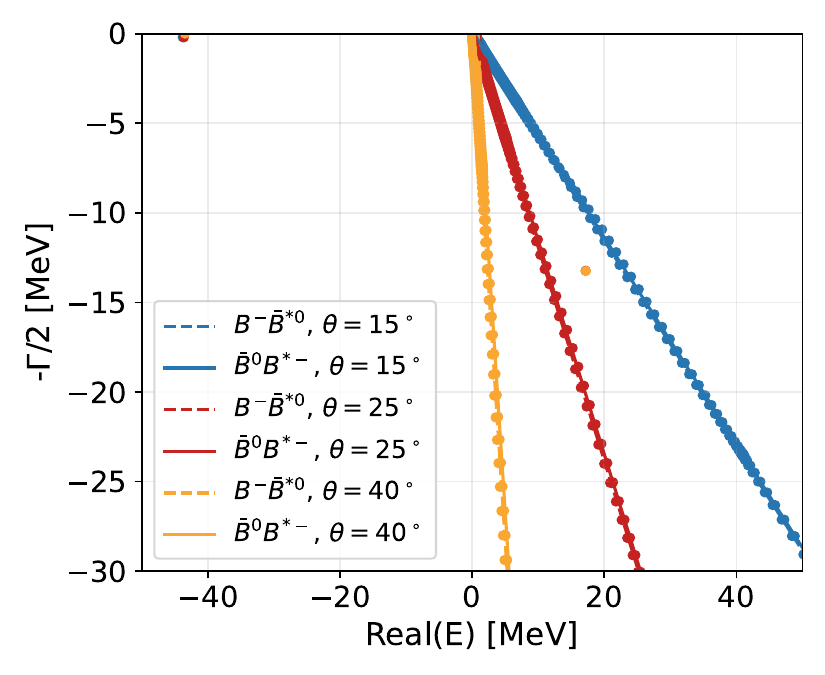}&
\includegraphics[width=0.5\textwidth]{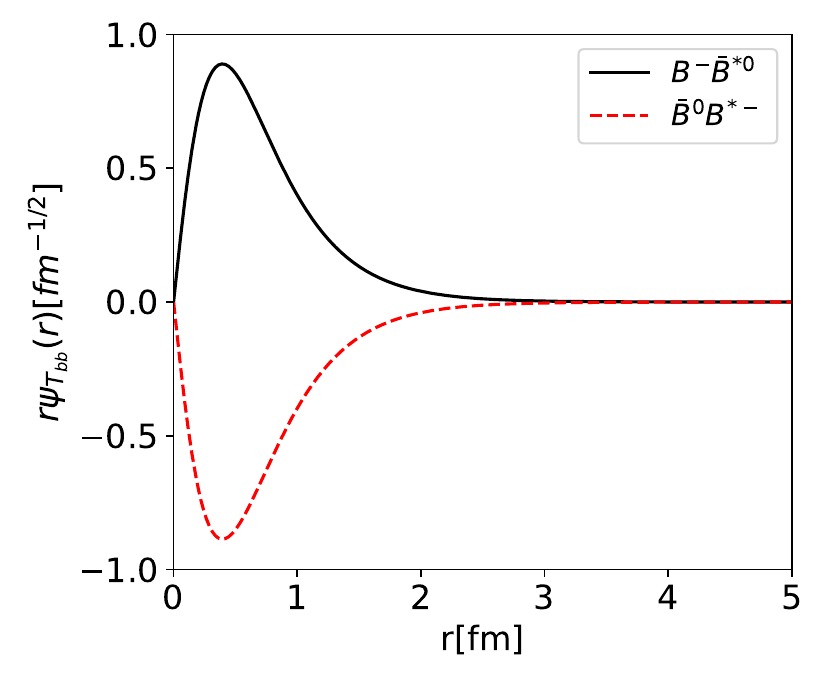}
\end{tabular}
\caption{The complex eigenvalues (left) and wave functions (right) of the two components with cutoff $\Lambda=1.0$ GeV for the $\bar{B}\bar{B}^*$ system.}\label{rtbb}
\end{figure}

To provide a direct perspective, the results using the CSM method with a cutoff $\Lambda=1.0$ GeV are presented in Fig.~\ref{rtbb}, and those for  $\Lambda=0.8$ and $1.2$ GeV are similar. We varied $\theta$ across three values to identify the stable eigenvalues corresponding to the bound and resonant states. The complex eigenvalues are displayed in the left panel of Fig.~\ref{rtbb}. Most of them lie along the continuum line and exhibit a rotation by an angle of $2\theta$, which corresponds to the scattering states. The two continuum lines almost overlap due to the small mass difference between $B^{-}\bar B^{*0}$ and $\bar B^0 B^{*-}$. Notably, one eigenvalue overlaps across three $\theta$ values and appears nearly stable on the real axis below the threshold. This shows that a bound state exists below $B^{-}\bar B^{*0}$ and $\bar B^0 B^{*-}$ thresholds with mass $M=10560$ MeV. In this state, the vector meson and pion exchange interactions, i.e. $V_{\rho}$ and $V_{\pi}$, significantly contribute to the formation of the deeply bound state. 

In the right panel of Fig.~\ref{rtbb}, we display the wave functions of $B^{-}\bar B^{*0}$ and $\bar B^0 B^{*-}$ within the bound state, which are similar but have opposite signs, clearly indicating the bound state to be an $I=0$ state as shown in Table~\ref{tab:obe}.
We also evaluated the properties of the bound state, including the root mean square radius, proportions of $B^{-}\bar B^{*0}$ and $\bar B^0 B^{*-}$ channels, and the ratio of the residue in two channels. These information obtained with three different cutoff values, is summarized in Table~\ref{tab:Tbb}. With the proportions,  the bound state is also shown to be almost purely an $I=0$ state indicating that isospin symmetry is well preserved due to the large mass of the bottom mesons.

In the left panel of Fig.~\ref{rtbb}, we also find a pole in the second RS, above the $B^{-}\bar B^{*0}$ and $\bar B^0B^{-}$ thresholds. As discussed in Sec. \ref{sec:csm}, the scaling angle $\theta$ must be larger than $\frac{1}{2} {tan}^{-1}(\Gamma/2 E_r)=18.8^\circ$ to identify the resonance pole $E_r-i\frac{\Gamma}{2}$. For $\theta=15^\circ< 18.8^\circ$, the resonant state cannot be normalizable by integration along the complex path $pe^{-i\theta}$ and cannot be identified. When $\theta>18.8^\circ$, such as $\theta=25^\circ$  and $\theta=40^\circ$, the resonant pole appears and exhibits overlap.  The pole positions for three different cutoff values are presented in Table~\ref{tab:resonance}. 

\begin{figure}[htp]
\centering
\begin{tabular}{ccc}
\includegraphics[width=0.5\textwidth]{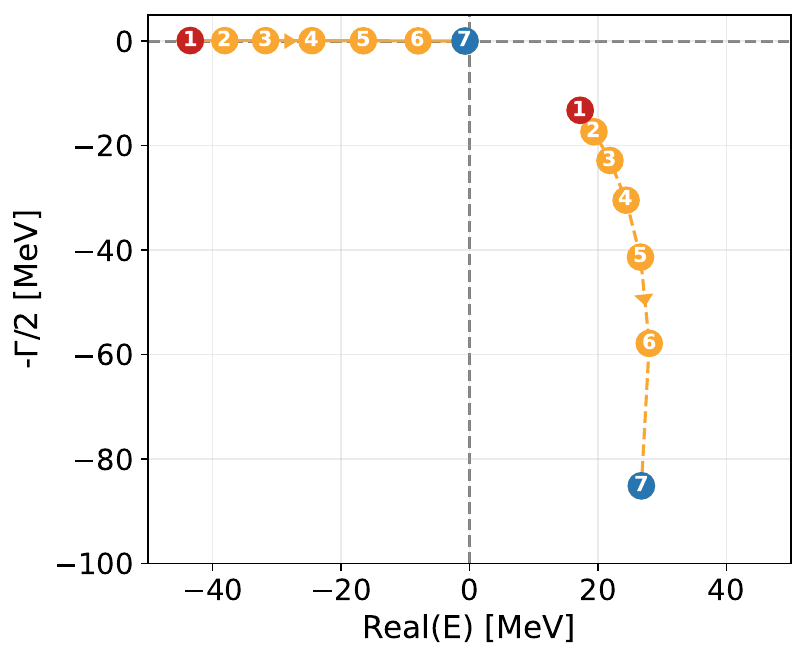}
\end{tabular}
\caption{The pole trajectories with varying meson mass parameters are shown. The circled numbers $1$ to $7$ represent sequential tuning of the component meson masses from  $B^{(*)}$ mesons to $D^{(*)}$ mesons. Red circles indicate resonance and bound states in the $\bar{B}\bar{B}^*$ system, while blue circles indicate those in the $DD^{*}$ system.} \label{fig:BB-DD}
\end{figure}

Analysis of the wave functions shows that this resonant state is an $I=1$ state. The $u$-channel interactions $V^u_\pi$ and $V^u_{\rho/\omega}$ contributes attractively, although the $V^u_\pi$ is halved. The $t$-channel contributes repulsively. These dynamics result in a less attractive potential compared to the $I=0$ case, leading to the formation of resonant states rather than virtual or bound states.   
%
%
%
%
An analogous state in the charm sector appears far from the real axis at $E(\Gamma)=3902.4(170.3)$ MeV, making it challenging to be observed in experiment due to its large decay width. Compared to the charm mesons, the much heavier mass of bottom mesons makes it easier for them to bind together. As a result, the binding energy in the $I=0$ sector  is much deeper, and a resonance  with a small decay width is generated in $I=1$ sector, both of which differ from the $T_{cc}$ case. To clarify this, we also illustrate the shift in pole positions as the meson masses are tuned down from those of the $B^{(*)}$ in Fig.~\ref{fig:BB-DD}. The results show that with larger meson masses, the bound states become more deeply bound, while the resonances approach the origin with a narrower decay width.


Note that the masses and widths of the bound state and resonance, extracted using the CSM and $T$-matrix,  exhibit a high degree of concurrence, as shown in Tables~\ref{tab:Tbb} and \ref{tab:resonance}.
This is because the Schr\"{o}dinger equation is essentially equivalent to the Lippmann-Schwinger equation, with the eigenvalues of the former corresponding to the poles of the $T$-matrix in the latter.
By altering the integral path in the complex scaling, the bound and resonant states solutions will not shift as long as the rotation avoids the singularity of the effective potential ${\cal V}$  and does not alter its damping behavior at infinity. 
The utilization of both algorithms could mutually supplement each other, enhancing the reliability of the numerical results. 
%

    \begin{table}
        \centering
        \linespread{1.3}
        \caption{The properties of the bound state in the $\bar{B}\bar{B}^*$ system with three cutoff values 
                 $\Lambda=0.8$, $1.0$, and $1.2$ GeV. The script ``BE" denotes the binding energy and ``$P$" stands for the proportion of the channel. 
                 The ratio of the residue in two channels is listed in the last column.}
        \label{tab:Tbb}
        \begin{tabular}{p{1.5cm}<{\centering} | p{2.3cm}<{\centering} | p{2.2cm}<{\centering} p{1.8cm}<{\centering} 
                        p{1.5cm}<{\centering} p{1.8cm}<{\centering} p{1.8cm}<{\centering} p{2.3cm}<{\centering}}  \hline\hline
        \multirow{2}{*}{$\Lambda$ (GeV)} &  $T$-matrix & \multicolumn{6}{c}{CSM} \\
            \cline{2-2}
            \cline{3-8}
            & Mass (MeV)  & Mass (MeV) & BE(MeV) &  $\sqrt{\langle r^2\rangle}$ & $P(B^{-}\bar B^{*0})$ & $P(\bar B^0B^{*-})$ & $|\frac{\text{Res}(\bar B^0B^{*-})}{\text{Res}(B^{-}\bar B^{*0})}|$ \\ \hline
            $0.8$   & $10572.2$  & $10572.2$  & 31.8 &  $0.74$ fm & $50.3\%$  & $49.7\%$  & $0.994$ \\
            $1.0$   & $10560.1$  & $10560.1$  & 43.9 & $0.61$ fm & $50.2\%$  & $49.8\%$  & $0.997$ \\
            $1.2$   & $10542.0$  & $10542.0$  & 62.0 & $0.51$ fm & $50.2\%$  & $49.8\%$  & $0.998$ \\ \hline\hline
        \end{tabular}
    \end{table}
    \begin{table}
    \caption{The properties of the resonance pole in the $\bar{B}\bar{B}^*$ systems with the cutoff values $\Lambda=0.8$, $1.0$, and $1.2$ GeV. }
    \label{tab:resonance}
    \begin{tabular}{p{2.5cm}<{\centering}|p{3.2cm}<{\centering}|p{3.2cm}<{\centering}|
                    p{3.2cm}<{\centering}|p{3.2cm}<{\centering}}
        \hline \hline
        \multirow{2}{*}{$\Lambda$ (GeV)} & \multicolumn{2}{c|}{$T$-matrix} & \multicolumn{2}{c}{CSM} \\
        \cline{2-3}
        \cline{4-5}
              & Mass (MeV) & Width (MeV) & Mass (MeV) & Width (MeV) \\ \hline
        $0.8$ & $10621.3$    & $20.6$      & $10621.3$    & $20.8$ \\
        $1.0$ & $10621.2$    & $26.4$      & $10621.2$    & $26.4$ \\
        $1.2$ & $10610.9$    & $17.0$      & $10610.9$    & $17.4$ \\
        \hline \hline
    \end{tabular}
\end{table}
{\subsection{$\bar{B}^*\bar{B}^*$ system}\label{sec:1chan2}

The $I(J^P)$ quantum numbers of $S$-wave $\bar{B}^*\bar{B}^*$ systems are constrained to be $I(J^P)=0(1^+)$,  $1(0^+)$  or $1(2^+)$. In this work, using the CSM, we explore the $\bar{B}^*\bar{B}^*$ system with $I(J^P)=0(1^+)$ and find a bound state.

In Table~\ref{tab:B*B*}, we present the numerical results from the CSM method for different cutoff values.
For the cutoff $\Lambda=1.0$ GeV,  the complex eigenvalues and wave function for the $\bar{B}^{*} \bar{B}^{*}$ system are shown in Fig.~\ref{rtbsbs}. As indicated in Table~\ref{tab:Tbb} and Table~\ref{tab:B*B*}, the binding energies and root-mean-square radii are nearly identical for the $\bar B \bar B^*$ and $\bar B^*\bar B^* $ bound states. This striking similarity is attributed to heavy quark spin symmetry whereby the $\bar B \bar B^*$ and $\bar B^*\bar B^* $ systems have the same potentials in the heavy quark limit. The existence of the $\bar B^*\bar B^* $ bound state has also been predicted using various model, such as the Weinberg’s scheme~\cite{Wang:2018atz}, the constituent interchange model~\cite{Yu:2019sxx}, the local hidden gauge approach~\cite{Dai:2022ulk}, the color-magnetic interaction~\cite{Luo:2017eub} and the OBE model~\cite{Ohkoda:2012hv,Ding:2020dio}.
\begin{figure}[htp]
\centering
\begin{tabular}{ccc}
\includegraphics[width=0.5\textwidth]{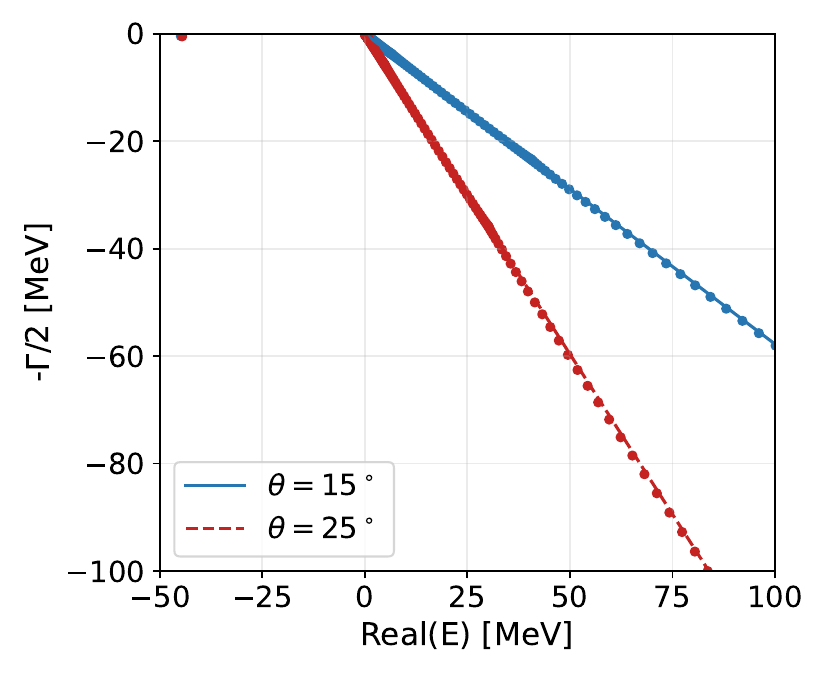}&
\includegraphics[width=0.5\textwidth]{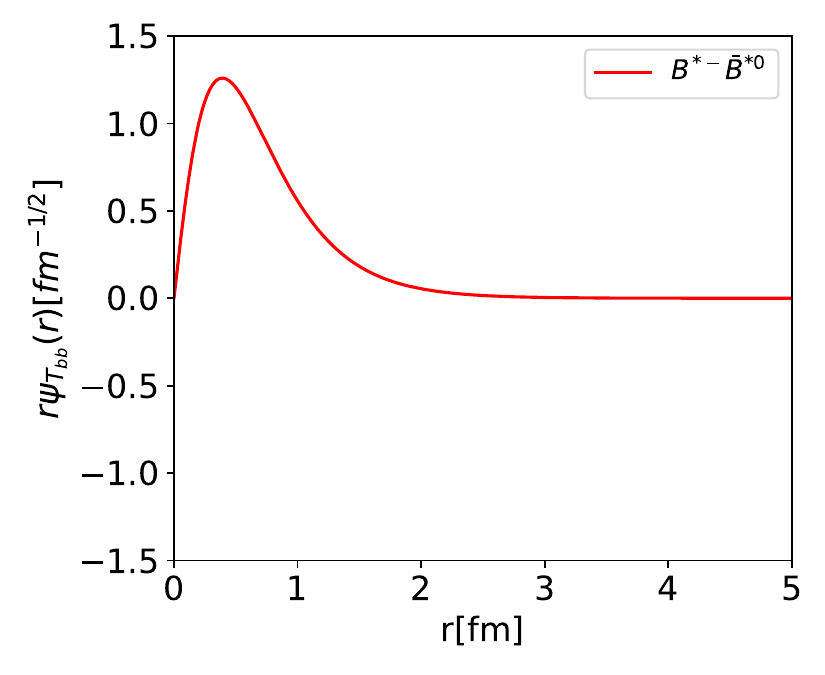}
\end{tabular}
\caption{The complex eigenvalues (left) and wave functions (right) of the component with cutoff $\Lambda=1.0$ GeV for the $\bar{B}^*\bar{B}^*$ system.}\label{rtbsbs}
\end{figure}
    \begin{table}
        \centering
        \linespread{1.3}
        \caption{The properties of the bound state in the $\bar{B}^*\bar{B}^*$ systems with three cutoff values $\Lambda=0.8$, $1.0$, and $1.2$ GeV. The script ``BE" denotes the binding energy.  }
        \label{tab:B*B*}
	  \begin{ruledtabular}
		\begin{tabular}{c|ccccccc}
			$ \quad \quad \Lambda$ (GeV)  \qquad \qquad  & Mass (MeV) & BE (MeV) & $\sqrt{\langle r^2\rangle}$ \\ \hline
			$0.8$ & $10616.8$ & 32.6 & $0.73$ fm  \\
			$1.0$ & $10604.3$ & 45.1 & $0.60$ fm  \\
			$1.2$ & $10585.5$ & 63.9 & $0.50$ fm  \\
		\end{tabular}
	\end{ruledtabular}
    \end{table}
\begin{table}[tp]
	\caption{The mass comparison of the $I(J^P) = 0(1^+)$ bound state in the doubly-bottom $\bar{b}\bar{b}ud$ tetraquark with previous theoretical studies and lattice QCD results is conducted. 
 }
        \label{tab:pre}
		\begin{tabular}{c|c} \hline\hline
			                                   & Mass (MeV) \\ \hline
                chiral constituent quark model    & $10261$~\cite{Vijande:2003ki} \\
                relativistic potential model     & $10504/10519$~\cite{Janc:2004qn} \\
                dynamical quark model             & $10426$~\cite{Vijande:2006jf} \\
                relativistic quark model          & $10502$~\cite{Ebert:2007rn} \\
                chiral SU($3$) quark model        & $10576$~\cite{Zhang:2007mu}, $10590$~\cite{Chen:2021tnn} \\ 
                nonrelativistic quark model       & $10560.3\sim10603.8$~\cite{Deng:2021gnb} \\                   
         strong and electromagnetic interactions  & $10389(12)$~\cite{Karliner:2017qjm} \\
                heavy-quark symmetry              & $10482$~\cite{Eichten:2017ffp} \\ 
                heavy diquark-antiquark symmetry  & $10488$~\cite{Cheng:2020wxa} \\ 
                lattice non-relativistic QCD      & $10415(10)(3)$~\cite{Francis:2016hui}, $10439(33)$~\cite{Junnarkar:2018twb}\\
                                                  & $10476(24)(10)$~\cite{Leskovec:2019ioa},  $10418(7)$~\cite{Mohanta:2020eed}, $10492(13)$~\cite{Hudspith:2023loy} \\ 
                QCD sum rule                      & $10200(300)$~\cite{Du:2012wp}, $10200(300)$~\cite{Navarra:2007yw}\\
                                                  & $10035(260)$~\cite{Agaev:2018khe}, $10135(240)$~\cite{Agaev:2020dba}\\                   
                non-relativistic quark model      & $10525$~\cite{Silvestre-Brac:1993zem}, $10482\sim10514$~\cite{Semay:1994ht} \\
                constituent quark model + OPE     & $10466.4$~\cite{Meng:2023for} \\
                four-body contact interactions + OPE & $10586.1(^{+10.3}_{-12.9})$~\cite{Wang:2018atz} \\
                the constituent interchange model & $10600.9$~\cite{Yu:2019sxx}\\
                one-meson exchange potential model& $10598(^{+2}_{-3})$~\cite{Ren:2021dsi}\\
                OBE                               & $10586.7$~\cite{Ke:2021rxd}, $10585.9$~\cite{Ding:2020dio} \\ 
                local hidden gauge approach       & $10583$~\cite{Dai:2022ulk}, $10583$~\cite{Feijoo:2023sfe} \\
                Born-Oppenheimer approximation    & $10522/10552$~\cite{Maiani:2022qze} \\             
                lattice QCD                       
                  & $10476 \pm 34 $~\cite{Leskovec:2019ioa}, $10514\pm 43$~\cite{Bicudo:2012qt,Bicudo:2015vta,Bicudo:2015kna,Bicudo:2017szl,Bicudo:2021qxj} \\ 
                  & $10415\pm 13 $~\cite{Francis:2016hui}, $10461 \pm 34 $~\cite{Junnarkar:2018twb}, $10449(17)$~\cite{Aoki:2023nzp} \\              
                  & $\sim 10484$~ \cite{Colquhoun:2022dte}, $\sim 10491$ ~\cite{Francis:2018jyb,Hudspith:2020tdf}, $10501 \pm 8$~\cite{Wagner:2022bff,Pflaumer:2022lgp}
                \\\hline 
                this work                         
                & $10542\sim10572$ ($\bar{B}\bar{B}^*$)\\
                & $10585\sim10616$ ($\bar{B}^*\bar{B}^*$)\\
                & $10380\sim10534$ ($\bar{B}\bar{B}^*$-$\bar{B}^*\bar{B}^*$)\\

                \hline\hline
		\end{tabular}
\end{table}
{\subsection{$\bar{B}\bar{B}^*$-$\bar{B}^*\bar{B}^*$ coupled channel effect}\label{sec:2chan}}
Given the close proximity in mass between $\bar B\bar B^*$  and $\bar B^*\bar B^*$ channels, it is imperative to consider the coupled channel effect. However, when this effect is taken into account, the binding energies become unexpectedly large, ranging from $127$ MeV to $319$ MeV with the cutoff $\Lambda$ in the range $[0.8,1.2]$ GeV and the energy shift due to the couppled channel effect is sensitive to the cutoff value. This puzzling phenomenon is discussed in Ref.~\cite{Nieves:2012tt}, where the authors used a sharp cutoff regulator to illustrate that the energy divergence after considering the coupled channel effect is at least proportional to $\Lambda$ or higher. This is because the coupled channel effect introduces cutoff dependence, associated with divergence, that should in principle be absorbed by the coupling constant. However, it is not expected that such dependence can be absorbed by the potential without the coupled channel effect, which is already determined by reproducing the binding energy of $T_{cc}^+$. To soften the dependence,  it is necessary to carefully consider the parameter value adjustments after incorporating the coupled channel effect.

Ref.~\cite{Wang:2022mxy} proposed to use different cutoff values for the involved channels. In this work, we employ this method by choosing different cutoff values for the $\bar{B}^*\bar{B}^*$ and $\bar{B}\bar{B}^*$ channels, as shown in Table \ref{tab:tbb2}.  With this treatment, one bound state and two resonant states are identified as summarized in Table~\ref{tab:tbb2}-\ref{tab:resonance2}. The complex eigenvalue and wave functions of the bound state with a cutoff $\Lambda=1.0$ GeV are displayed in Fig.~\ref{rtbb2}. In Table~\ref{tab:pre}, we present the obtained mass of the bound state alongside results from other phenomenological studies and lattice QCD simulations for comparison.  

Comparing these results with those obtained without the coupled-channel effect (see Table~\ref{tab:Tbb} and Table~\ref{tab:resonance}),  it is evident that the coupled-channel effect plays a distinct and indispensable role in $\bar B \bar{B}^{*}$ system. Taking the cutoff $\Lambda=1.0$ GeV as an example, the binding energy of the deeply bound state increases from $44$ to $128$ MeV. The $\bar B^*\bar B^*$ component significantly mixes with  $\bar B\bar B^*$, comprising about $33\%$. The coupled-channel effect contributes to an effectively repulsive potential in the  $\bar B^*\bar B^*$ system. Consequently, the  $\bar B^*\bar B^*$ bound state, which appears  without the coupled-channel effect, transforms into a resonant state with a substantial decay width into the  $\bar B \bar B^*$ channels. The $\bar B\bar B^*$ resonance pole at $10621$ MeV remains nearly unchanged. As discussed above, this state is primarily an  $I=1$ state with minimal isospin-breaking effects. Therefore, it almost decouple with  the $\bar B^{*}\bar{B}^{*}$ state with $I(J^P)=0(1^+)$.

In addition to the contribution from the $S$-wave, $D$-wave interaction also needed to be considered for $\bar{B}\bar{B}^*$-$\bar{B}^*\bar{B}^*$ coupled channel. Nevertheless, even though the coupled channel effect will deepen the bound $T_{bb}$ states importantly, it will not qualitatively change the existence of the bound state, nor will it influence the resonant $T_{bb}$ state. More experiments and lattice investigations are essential for constraining the parameters and further detailed discussions are warranted in the future research~\cite{Ren2024}. 

    \begin{table}
        \centering
        \linespread{1.3}
        \caption{The properties of the bound state in the $\bar{B}\bar{B}^*$-$\bar{B}^*\bar{B}^*$ coupled channel system with three sets of cutoff values. $\Lambda$ and $\Lambda^{\prime}$ denote the cutoff parameters of the $\bar{B}\bar{B}^*$ and $\bar{B}^*\bar{B}^*$ channels, respectively. The script ``BE" denotes the binding energy. The script ``$P$" stands for the proportion of the channel. }
        \label{tab:tbb2}
	  \begin{ruledtabular}
		\begin{tabular}{cc|ccccccc}
			$ \Lambda$ (GeV) & $\Lambda^{\prime}$ (GeV) & Mass (MeV) & BE (MeV) & $\sqrt{\langle r^2\rangle}$ & $P(B^{-}\bar B^{*0})$ & $P(\bar B^0B^{*-})$&
			$P(\bar B^{*0}B^{*-})$\\ \hline
			$0.8$ & $0.6$ & 10533.5 & 70.6  & 0.62 & 38.2 & 38.0 & 23.8 \\
			$1.0$ & $0.8$ & 10476.4 & 127.7 & 0.49 & 33.8 & 33.7 & 32.5 \\
			$1.2$ & $1.0$ & 10380.2 & 223.8 & 0.40 & 30.9 & 30.8 & 38.3 \\
		\end{tabular}
	\end{ruledtabular}
    \end{table}
\begin{figure}[htp]
\centering
\begin{tabular}{ccc}
\includegraphics[width=0.5\textwidth]{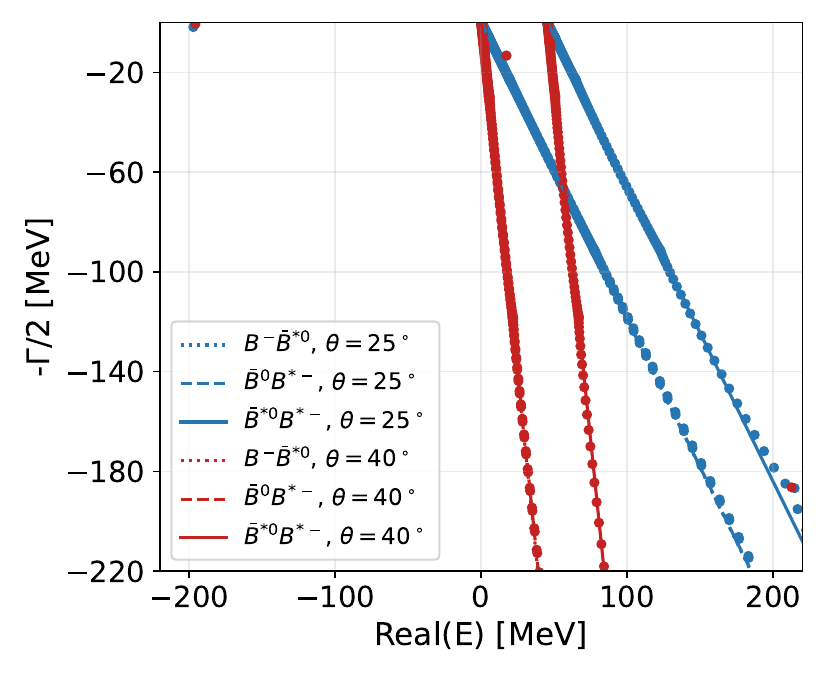}&
\includegraphics[width=0.5\textwidth]{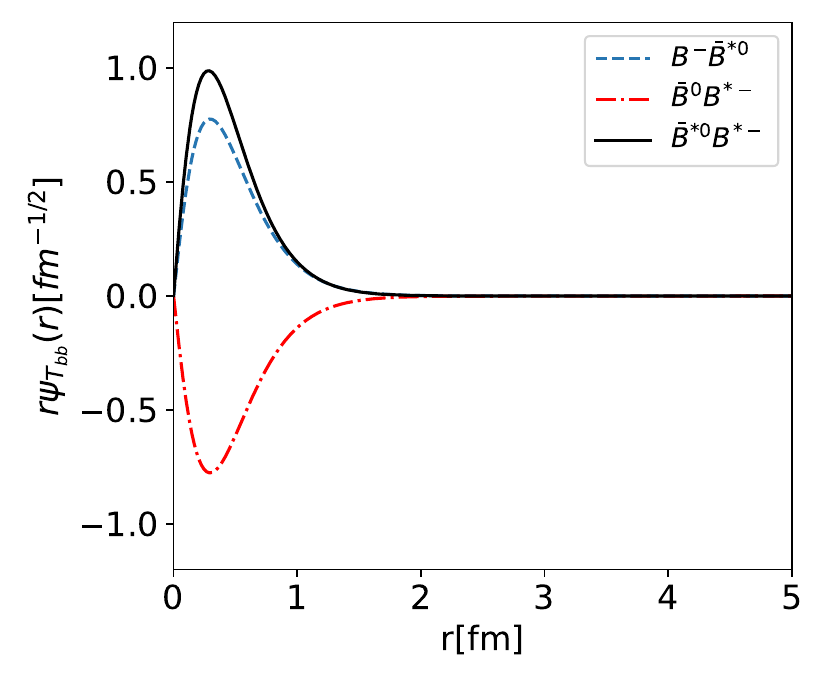}
\end{tabular}
\caption{The complex eigenvalues (left) and wave functions (right) of the components with cutoff $\Lambda=1.0$ GeV for the $\bar{B} \bar{B}^{*}$-$\bar{B}^{*} \bar{B}^{*}$ system without the delta potential.}\label{rtbb2}
\end{figure}
\begin{table}
    \centering
    \caption{The properties of the resonance pole in the $\bar{B}\bar{B}^*$-$\bar{B}^*\bar{B}^*$ coupled channel system with three sets of cutoff values. $\Lambda$ and $\Lambda^{\prime}$ denote the cutoff parameters of the $\bar{B}\bar{B}^*$ and $\bar{B}^*\bar{B}^*$ channels, respectively.}
    \label{tab:resonance2}
    \begin{tabular}
        {p{2.15cm}<{\centering} p{2.15cm}<{\centering}|p{2.8cm}<{\centering}|p{2.8cm}<{\centering}|p{2.8cm}<{\centering}|p{2.8cm}<{\centering}} \hline \hline
        \multirow{2}{*}{$\Lambda$ (GeV)} & \multirow{2}{*}{$\Lambda^{\prime}$ (GeV)} & \multicolumn{2}{c|}{Resonance \uppercase\expandafter{\romannumeral1}} & \multicolumn{2}{c}{Resonance \uppercase\expandafter{\romannumeral2}} \\
        \cline{3-6}
              &       & Mass (MeV) & Width (MeV) & Mass (MeV) & Width (MeV) \\ \hline
        $0.8$ & $0.6$ & $10621.3$  & $20.8$      & $10700.5$  & $212.7$ \\
        $1.0$ & $0.8$ & $10621.2$  & $26.5$      & $10755.7$  & $308.3$ \\
        $1.2$ & $1.0$ & $10611.0$  & $17.3$      & $10847.6$  & $409.5$ \\
        \hline \hline
    \end{tabular}
\end{table}
\begin{figure}[htp]
\centering
\begin{tabular}{ccc}
\includegraphics[width=0.5\textwidth]{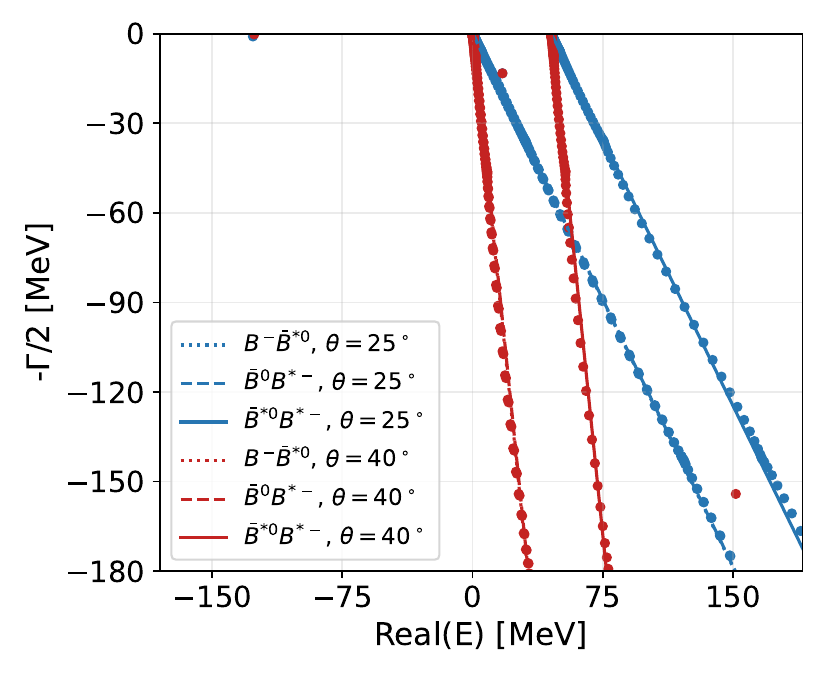}&
\includegraphics[width=0.5\textwidth]{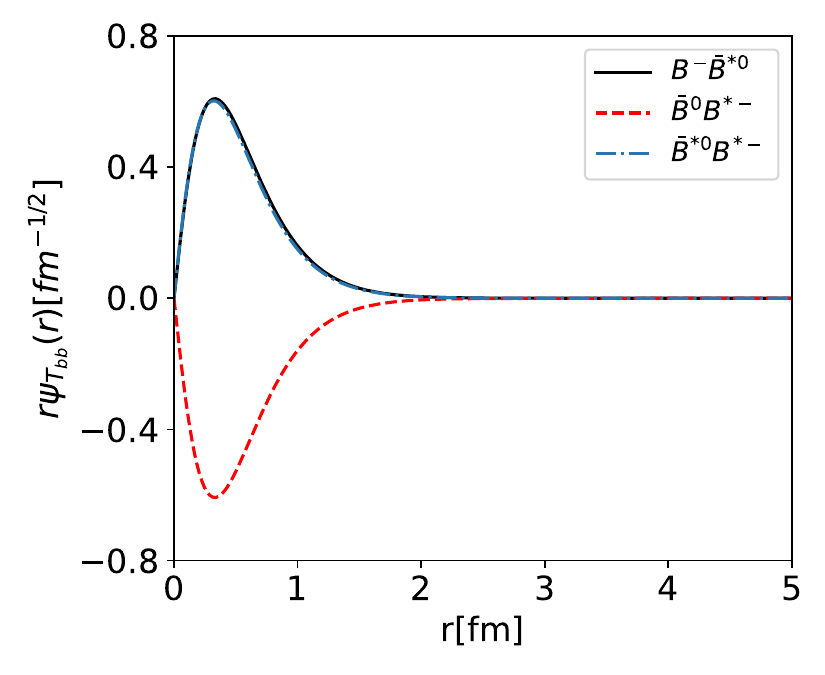}
\end{tabular}
\caption{The complex eigenvalues (left) and wave functions (right) of the components with cutoffs $\Lambda=1.0$ GeV and $\Lambda^{\prime}=0.8$ GeV for the $\bar{B} \bar{B}^{*}$-$\bar{B}^{*} \bar{B}^{*}$ system without the delta potential.}\label{rtbb4}
\end{figure}
\vspace{10pt}
\section{Summary}\label{sec:sum}
In this work, we searched for possible $\bar{B}^{(*)}\bar{B}^*$ hadronic molecules with $J^P=1^+$. The meson-meson interaction occurring by the exchange of light mesons can be formulated based on the OBE model. Within the heavy quark symmetry, we adopted the same coupling constants and cutoff parameter as those used in the $DD^*$ case for the $T^+_{cc} $ state, as investigated in our previous paper. We identified  dynamically generated states utilizing $T$-matrix pole analysis and CSM. The results from these two methods are remarkably consistent.

For the $T_{bb}$ system, the isospin symmetry is well-preserved. A bound state with $I(J^P)=0({1^+})$ was identified, positioned approximately $44$ MeV below the $\bar{B}\bar{B}^*$ threshold. Its properties, including the root mean square radius and component proportion were determined. Additionally, the analysis revealed the presence of a resonant  state with $I(J^P)=1{(1^+)}$ at $10621.2$ MeV. For the  $\bar{B}^{*} \bar{B}^{*}$ system, the state with $J^{P}=1^+$  can only be an isoscalar state. It has the same potentials as  the  $I(J^P)=0({1^+})$ $\bar{B} \bar{B}^{*}$ state in the heavy quark limit. Thus, we also found evidence for the existence of a bound $I(J^P)=0({1^+})$ $\bar{B} \bar{B}^{*}$ state with a binding energy of $45.1$ MeV and a mass of $10604.3$ MeV.

After considering the $\bar{B}\bar{B}^*$-$\bar{B}^*\bar{B}^*$ coupled channel effect, the binding energy becomes substantially deeper, resulting in the distortion of the potential determined by the $T_{cc}$. 
Meanwhile, the $\bar{B}^*\bar{B}^*$ bound state will be transformed into a resonance.
When tuning the cutoff within a narrow range, we found a strong dependence of the results on the cutoff for the $T_{bb}$ system. 
Additional constrains should be introduced for this study of coupled channel effect in the future.
The information about the dynamical generated states in the $\bar{B}^{(*)}\bar{B}^*$ system presented in this work is anticipated to provide valuable assistance and guidance for future experimental endeavors aimed at exploring the existence of these states. 

\begin{acknowledgments}

This work is partly supported  by the KAKENHI under Grant Nos. 23K03427 and 24K17055 (G.J.W), and by the National Natural Science Foundation of China (NSFC) under Grants Nos.~12275046 (Z.Y.), 12175239 and 12221005 (J.J.W), and by the National Key R$\&$D Program of China under Contract No. 2020YFA0406400 (J.J.W), by the Xiaomi Foundation / Xiaomi Young Talents Program (J.J.W).

\end{acknowledgments}

\begin{appendix}

\section{Potentials associated with the $\bar{B}^{(*)}\bar{B}^{(*)}$ channel\label{app:vBB}}
With the procedure described in Subsec.~\ref{sec:eff}, the effective potentials in the momentum space for $\bar{B}\bar{B}^*$-$\bar{B}^*\bar{B}^*$ channels are derived using the Lagrangian in Eqs.~(\ref{Lagrangian:p})-(\ref{lagrangian:v}), as shown in this section.
The scattering process ${\mathcal{P}_1^{(*)}(p_1)\mathcal{P}_2^{(*)}(p_2)\to \mathcal{P}_3^{(*)}(p_3)\mathcal{P}_4^{(*)}(p_4)}$ involve $t$ and $u$ channel contribution,  as illustrated in Fig.~\ref{fei}, 
\begin{figure}[htp]
\centering
\begin{tabular}{ccc}
\includegraphics[width=0.43\textwidth]{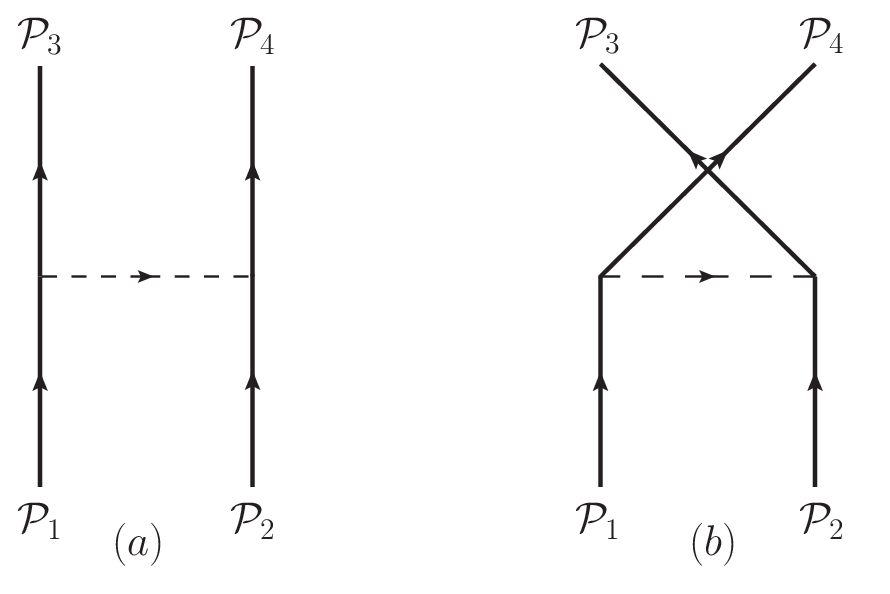}
\end{tabular}
\vspace{-15pt}
\caption{The Feynman diagrams for the $\bar{B}^{(*)} \bar{B}^{(*)}$ system, where $(a)$ represents direct channel and $(b)$ represents crossed channel. $\mathcal{P}$ corresponds to  $\bar{B}$ or $\bar{B}^{*}$ mesons.}\label{fei}
\end{figure}
\begin{itemize}
\item ${\mathcal{P}\mathcal{P}^{*}\to \mathcal{P}\mathcal{P}^{*}}$
\begin{eqnarray}
V_{\pi}^{u}&=&\frac{g^2}{f_\pi^2}\frac{(\epsilon_4^{\dagger}\cdot q)(q\cdot\epsilon_2)}{q^2-m_{\pi}^2},\label{eq:v1}\\
V_{\rho/\omega}^{u}&=&-2\lambda^2g_{V}^2\frac{(\epsilon_4^{\dagger}\cdot q)(q\cdot\epsilon_2)-q^2(\epsilon_4^{\dagger}\cdot\epsilon_2)}{q^2-m_{\rho/\omega}^2},\label{eq:v2}\\
V_{\rho/\omega}^{t}&=&\frac{\beta^2g_{V}^2}2\frac{(\epsilon_4^{\dagger}\cdot\epsilon_2)}{q^2-m_{\rho/\omega}^2},
\end{eqnarray}
\item ${\mathcal{P}\mathcal{P}^{*}\to \mathcal{P}^{*}\mathcal{P}^{*}}$
\begin{eqnarray}
V_{\pi}^{u}&=&\frac{g^2}{f_{\pi}^2}\frac{[\vec{q}\cdot i({\vec{\epsilon}}_{3}^{\dagger}\times\vec{\epsilon}_2)][\vec{\epsilon}_4^{\dagger}\cdot \vec{q}]}{\vec{q}^2-m_{\pi}^2},\\
V_{\pi}^{t}&=&\frac{g^2}{f_{\pi}^2}\frac{[\vec{q}\cdot i(\vec{\epsilon}_4^{\dagger}\times\vec{\epsilon}_2)][\vec{\epsilon}_3^{\dagger}\cdot \vec{q}]}{\vec{q}^2-m_{\pi}^2},\\
V_{\rho/\omega}^{u}&=&-2\lambda^2g_{V}^2\frac{[\vec{q}\cdot i(\vec{\epsilon}_4^{\dagger}\times\vec{\epsilon}_2)][\vec{\epsilon}_3^{\dagger}\cdot \vec{q}]-[\vec{q}\cdot i(\vec{\epsilon}_4^{\dagger}\times\vec{\epsilon}_3^{\dagger})][\vec{q}\cdot\vec{\epsilon}_2]}{\vec{q}^2-m_{\rho/\omega}^2},\\
V_{\rho/\omega}^{t}&=&-2\lambda^2g_{V}^2\frac{[\vec{q}\cdot i(\vec{\epsilon}_3^{\dagger}\times\vec{\epsilon}_2)][\vec{\epsilon}_4^{\dagger}\cdot \vec{q}]-[\vec{q}\cdot i(\vec{\epsilon}_3^{\dagger}\times\vec{\epsilon}_4^{\dagger})][\vec{q}\cdot\vec{\epsilon}_2]}{\vec{q}^2-m_{\rho/\omega}^2},
\end{eqnarray}
\item ${\mathcal{P}^{*}\mathcal{P}^{*}\to \mathcal{P}\mathcal{P}^{*}}$
\begin{eqnarray}
V_{\pi}^{u}&=&\frac{g^2}{f_{\pi}^2}\frac{[\vec{q}\cdot i(\vec{\epsilon}_4^{\dagger}\times\vec{\epsilon}_1)][\vec{\epsilon}_2\cdot \vec{q}]}{\vec{q}^2-m_{\pi}^2},\\
V_{\pi}^{t}&=&\frac{g^2}{f_{\pi}^2}\frac{[\vec{q}\cdot i(\vec{\epsilon}_4^{\dagger}\times\vec{\epsilon}_2)][\vec{\epsilon}_1\cdot \vec{q}]}{\vec{q}^2-m_{\pi}^2},\\
V_{\rho/\omega}^{u}&=&-2\lambda^2g_{V}^2\frac{[\vec{q}\cdot i(\vec{\epsilon}_4^{\dagger}\times\vec{\epsilon}_2)][\vec{q}\cdot\vec{\epsilon}_1]-[\vec{q}\cdot i(\vec{\epsilon}_1\times\vec{\epsilon}_2)][\vec{\epsilon}_4^{\dagger}\cdot \vec{q}]}{\vec{q}^2-m_{\rho/\omega}^2},\\
V_{\rho/\omega}^{t}&=&-2\lambda^2g_{V}^2\frac{[\vec{q}\cdot i(\vec{\epsilon}_4^{\dagger}\times\vec{\epsilon}_1)][\vec{q}\cdot\vec{\epsilon}_2]-[\vec{q}\cdot i(\vec{\epsilon}_2\times\vec{\epsilon}_1)][\vec{\epsilon}_4^{\dagger}\cdot \vec{q}]}{\vec{q}^2-m_{\rho/\omega}^2},
\end{eqnarray}
\item ${\mathcal{P}^{*}\mathcal{P}^{*}\to \mathcal{P}^{*}\mathcal{P}^{*}}$
\begin{eqnarray}
V_{\pi}^{u}&=&\frac{g^2}{f_\pi^2}\frac{[\vec{q}\cdot i(\vec{\epsilon}_4^{\dagger}\times\vec{\epsilon}_1)][\vec{q}\cdot i(\vec{\epsilon}_3^{\dagger}\times\vec{\epsilon}_2)]}{\vec{q}^2-m_{\pi}^2},\\
V_{\pi}^{t}&=&\frac{g^2}{f_\pi^2}\frac{[\vec{q}\cdot i(\vec{\epsilon}_3^{\dagger}\times\vec{\epsilon}_1)][\vec{q}\cdot i(\vec{\epsilon}_4^{\dagger}\times\vec{\epsilon}_2)]}{\vec{q}^2-m_{\pi}^2},\\
V_{\rho/\omega}^{u}(\lambda)&=&-2\lambda^2g_{V}^2\frac{[\vec{q}\cdot i(\vec{\epsilon}_4^{\dagger}\times\vec{\epsilon}_2)][\vec{q}\cdot i(\vec{\epsilon}_3^{\dagger}\times\vec{\epsilon}_1)]-\vec{q}^2 [i(\vec{\epsilon}_4^{\dagger}\times\vec{\epsilon}_1)][i(\vec{\epsilon}_3^{\dagger}\times\vec{\epsilon}_2)]}{\vec{q}^2-m_{\rho/\omega}^2},\\
V_{\rho/\omega}^{t}(\lambda)&=&-2\lambda^2g_{V}^2\frac{[\vec{q}\cdot i(\vec{\epsilon}_3^{\dagger}\times\vec{\epsilon}_1)][\vec{q}\cdot i(\vec{\epsilon}_4^{\dagger}\times\vec{\epsilon}_2)]-\vec{q}^2 [i(\vec{\epsilon}_3^{\dagger}\times\vec{\epsilon}_1)][i(\vec{\epsilon}_4^{\dagger}\times\vec{\epsilon}_2)]}{\vec{q}^2-m_{\rho/\omega}^2},\\
V_{\rho/\omega}^{u}(\beta)&=&\frac{\beta^2g_{V}^2}2\frac{[ i(\epsilon_4^{\dagger}\cdot\epsilon_1)][i(\epsilon_3^{\dagger}\cdot\epsilon_2)]}{q^2-m_{\rho/\omega}^2},\label{eq:v3}\\
V_{\rho/\omega}^{t}(\beta)&=&\frac{\beta^2g_{V}^2}2\frac{[ i(\epsilon_3^{\dagger}\cdot\epsilon_1)][i(\epsilon_4^{\dagger}\cdot\epsilon_2)]}{q^2-m_{\rho/\omega}^2},\label{eq:v3}
\end{eqnarray}
\end{itemize}
where $m_{\pi/\rho/\omega}$ represent the exchanging meson masses and $f_{\pi}=132$ MeV refers to the $\pi$ meson decay constant.  ${\epsilon_i}$ ($i=1,2,3,4$) represent the polarization vector of the vector meson $\mathcal{P}_i^{*}$. 
The constant $g=0.57$ is determined from the strong decay $D^*\rightarrow D\pi$  ~\cite{ParticleDataGroup:2022pth,CLEO:2001foe,CLEO:2001sxb}.
To compare our parameters with others from the phenomenological estimation~\cite{Li:2012ss,Li:2012cs,Wang:2018jlv}, we introduce a constant $g_V=5.8$ in the Lagrangians, as shown in Eqs. \eqref{Lagrangian:p} and \eqref{lagrangian:v}. 
\end{appendix}
\bibliography{ref.bib}
\end{document}